\documentclass[12pt]{article}

\usepackage[body={17.5cm, 21cm},right=2cm]{geometry}

\usepackage{color}
\usepackage{graphicx}
\usepackage{epsf}
\usepackage{graphicx,epsfig}
\usepackage{amsmath}
\usepackage{amssymb}
\usepackage{epsfig}
\usepackage{cite}
\usepackage{color,colordvi}
\newcommand{\be}{\begin{eqnarray}}
\newcommand{\ee}{\end{eqnarray}}
\newcommand{\bi}{\begin{itemize}}
\newcommand{\ei}{\end{itemize}}

\def\p{\partial}
\newcounter{hran}

\def\MSbar{\relax\ifmmode\overline{\rm MS}\else{$\overline{\rm MS}${ }}\fi}

\def\de{{\rm d}}

 \def\vx{\vec{ x}} 
\def\vk{\vec{k}}

\def\vb{\vec{b}}
\def\lg{\Big<}
\def\rg{\Big>}
\def\vx{\vec{x}}

\def\t{\tau}

 \def\vx{\vec{ x}} 
\def\vk{\vec{k}}

\numberwithin{equation}{section}
\begin{document}
\vspace{5mm}
\vspace{0.5cm}
\begin{center}

\def\thefootnote{\fnsymbol{footnote}}

{\Large \bf 
Symmetries of Vector Perturbations \\
\vspace{0.25cm}	
during the de Sitter Epoch }
\\[1.5cm]
{\large  M. Biagetti$^{a}$, A. Kehagias$^{b}$, E. Morgante$^{a}$, H. Perrier$^{a}$   and A. Riotto$^{a}$}
\\[0.5cm]

\vspace{.3cm}
{\normalsize { \it $^{a}$ Department of Theoretical Physics and Center for Astroparticle Physics (CAP)\\ 24 quai E. Ansermet, CH-1211 Geneva 4, Switzerland}}\\
\vspace{.3cm}
{\normalsize {\it  $^{b}$ Physics Division, National Technical University of Athens, \\15780 Zografou Campus, Athens, Greece}}\\

\vspace{.3cm}

\vspace{.3cm}
%{\normalsize { E-mail: kehagias@central.ntua.gr and antonio.riotto@unige.ch}}

%\vspace{.2cm}

\end{center}

\vspace{2cm}

\hrule \vspace{0.3cm}
{\small  \noindent \textbf{Abstract} \\[0.3cm]
\noindent 
We analyze  the class of models  where a 
 suitable coupling between the inflaton field and the vector field gives rise to  scale-invariant vector perturbations.
We exploit  the fact that   the de Sitter isometry group acts  as 
conformal group  on the three-dimensional Euclidean space  for  the super-Hubble fluctuations   
 in order to characterize the correlators involving the inflaton and the vector fields.

\vspace{0.5cm}  \hrule
\vskip 1cm

\def\thefootnote{\arabic{footnote}}
\setcounter{footnote}{0}
%\maketitle

%\date{\today}

\baselineskip= 15pt

\newpage 
%\tableofcontents

%%%%%%%%%%%%%%%%%%%%%%%%%%%%%%%%%%%%%%%%%%%%%%%%%%%%%%%%%%%%%%%%%%%%%%%%%%%%%
%%%%%%%%%%%%%%%%%%%%%%%%%%%%%%%%%%%%%%%%%%%%%%%%%%%%%%%%%%%%%%%%%%%%%%%%%%%%%

\section{Introduction}

%%%%%%%%%%%%%%%%%%%%%%%%%%%%%%%%%%%%%%%%%%%%%%%%%%%%%%%%%%%%%%%%%%%%%%%%%%%%%
%%%%%%%%%%%%%%%%%%%%%%%%%%%%%%%%%%%%%%%%%%%%%%%%%%%%%%%%%%%%%%%%%%%%%%%%%%%%%

\noindent
It has recently become  clear that 
symmetries play a crucial role    in  characterizing  the properties of  the  cosmological perturbations  generated by an  inflationary stage \cite{lrreview}.  During inflation  the de Sitter isometry group acts  as conformal group  on $\mathbb{R}^3$ when the fluctuations are on 
super-Hubble scales. During such a stage, correlators are constrained by conformal invariance as  the SO(1,4) isometry
of the de Sitter background is realized as conformal symmetry of the flat $\mathbb{R}^3$ sections 
\cite{antoniadis,pimentel,creminelli1,us1,us2}. This  happens when the  cosmological perturbations are sourced
by light scalar fields other than the inflaton (the field that drives inflation). In the opposite case in which the inflationary perturbations
originate from  only one degree of freedom,   conformal consistency relations among the inflationary correlators
 have also been recently  investigated  \cite{creminelli2,hui,baumann1,baumann2,nicolis,hui2}.  Similarly, one may study  the symmetries enjoyed by the Newtonian equations of motion of the non-relativistic dark matter fluid coupled to gravity which give rise to the phenomenon of gravitational instability  and reveal   consistency relations involving the soft limit of the $(n + 1)$-correlator functions of matter and galaxy overdensities \cite{krls,ppls}.

On the other hand, there has been recently a lot of interest in models which can produce vector field perturbations during inflation. There are mainly two reasons. On one side, one might hope to generate large-scale magnetic fields if vector field perturbations are excited during a de Sitter stage \cite{ruth}; on the other side claims of broken statistical invariance  of the CMB modes, as hinted also by the recent Planck satellite data  \cite{planck}, have put forward the proposal that such a breaking might be due to vector fields \cite{review vector, Yokoyama:2008xw,v0,v1,v2,v3,v4,v5,v6,v7}.

In this paper we shall investigate the symmetry properties of the vector field models with a kinetic term given by

\be
\label{wer}
{\cal L}=-\frac{1}{4}I^2(\phi) F_{\mu\nu}^2,
\ee
where $\phi$ indicates the inflaton field. 
Vector perturbations can be generated if the function $I(\phi)$ has the appropriate time dependence \cite{I0,I1}. In particular, if $I\sim a^n$, being $a$ the scale factor, magnetic modes are generated during inflation with a scale-invariant spectrum for $n=2$ and $n=-3$. In the first case, however, a too large electromagnetic coupling constant is generated during inflation \cite{I3,I4}, while the second case implies
a too large  energy density in the electric modes. Nevertheless, some recent work have investigated   the  cross-correlations
between primordial perturbations and large-scale magnetic fields induced by the coupling (\ref{wer})  
\cite{I5,motta,sloth,deltaN,lyth} as well as the (possibly too large) contribution from the vector modes 
to the anisotropic power spectrum of the curvature perturbation \cite{Gumrukcuoglu:2010yc, Dulaney:2010sq, Watanabe:2010fh,peloso,komatsu}.
For these reasons, the reader should be aware that it might not be healthy to  identify the vector field with  the electromagnetic field.
At any rate, the goal of this paper is to analyze the conformal symmetries enjoyed by the vector perturbations on super-Hubble scales. We will see that the action associated to the Lagrangian (\ref{wer}) respects the  conformal group  on $\mathbb{R}^3$ when the fluctuations are on 
super-Hubble scales and therefore the correlators involving the inflaton and the vector fields must be invariant
under conformal transformations of Euclidean three-space on the future boundary. This  will allow us  to write down the appropriate Ward identites as well as the two- and three-point correlator between the inflaton field and the vector fields, thus explaining some features found recently in the literature. 

The paper is organized as follows. In section 2, as a warm-up, we describe the conformal symmetries of de Sitter for the action of a massive scalar field; in section 3 we analyze the conformal symmetries of de Sitter in the presence of a vector field, and describe the correlators in section 4. Finally, section 5 contains our conclusions.

%%%%%%%%%%%%%%%%%%%%%%%%%%%%%%%%%%%%%%%%%%%%%%%%%%%%%%%%%%%%%%%%%%%%%%%%%%%%%
%%%%%%%%%%%%%%%%%%%%%%%%%%%%%%%%%%%%%%%%%%%%%%%%%%%%%%%%%%%%%%%%%%%%%%%%%%%%%

\section{Conformal symmetries of de Sitter and the scalar field}

%%%%%%%%%%%%%%%%%%%%%%%%%%%%%%%%%%%%%%%%%%%%%%%%%%%%%%%%%%%%%%%%%%%%%%%%%%%%%
%%%%%%%%%%%%%%%%%%%%%%%%%%%%%%%%%%%%%%%%%%%%%%%%%%%%%%%%%%%%%%%%%%%%%%%%%%%%%

Let us start by recalling some of the properties of the conformal symmetry in de Sitter. 
Conformal invariance 
in three-dimensional space $\mathbb{R}^3$ is connected to the symmetry under the group $SO(1,4)$ in the same way 
conformal invariance in a four-dimensional Minkowski spacetime is connected to the $SO(2,4)$ group. As $SO(1,4)$ is the 
isometry group of de Sitter spacetime,  a conformal phase during which fluctuations were generated could be 
a  de Sitter stage.  In such a  case, the kinematics  is specified by the embedding 
of $\mathbb{R}^3$ as flat sections in de Sitter spacetime. The de Sitter isometry group acts  as conformal group 
on $\mathbb{R}^3$ when the fluctuations are super-Hubble. It is in this regime that  the $SO(1,4)$ isometry
of the de Sitter background is realized as conformal symmetry of the flat $\mathbb{R}^3$ sections.  Correlators are expected to be constrained by conformal invariance. All these reasonings apply in the case in which the cosmological perturbations are generated 
by light scalar fields other than the inflaton, in particular vector perturbations. Indeed, it is only in such a case that correlators
inherit all the isometries of de Sitter.

Let us first describe the case of the scalar field. The de Sitter space in conformally flat coordinates is described by the  metric 
\be
 {\rm d}s^2=\frac{1}{H^2\tau^2}\left(-{\rm d}\tau^2+{\rm d} \vx^2\right). \label{met}
\ee 
It  can easily be checked that the transformations 
\be
&& x\to x_i'=a_i+M_{\,\,i}^jx_j, \\
&& x_i\to x_i'=\lambda x_i,~~~\tau\to \tau'=\lambda \tau, \\
&&x_i\to x_i'=\frac{x_i+b_i(-\tau^2+\vx^2)}{1+2\vec{b}\cdot\vec{x}+b^2(-\tau^2+\vx^2)}, \label{scalings}~~~
\tau\to \tau'=\frac{\tau}{1+2\vec{b}\cdot\vec{x}+b^2(-\tau^2+\vx^2)}, \label{specconf}
\ee 
are isometries of the de Sitter metric. 
 They correspond to translations (by a vector $\vec{a}$), rotations ($M^{i}_{\,\,j}$), dilations
(by a real parameter $\lambda$) and special conformal transformations (parametrized by a real vector $\vec{b}$), respectively.  
 In particular, for infinitesimal parameters and for super-Hubble scales ($\tau\to 0$), the special conformal transformations read
 
 \be
 \label{asdf}
 \vx\to  \vx'=\vx+\delta\vx=\vx(1-2\vec{b}\cdot\vec{x})+\vb\,\vx^2,\,\,\,\,
\tau\to \tau'=\tau+\delta\tau=\tau(1-2\vec{b}\cdot\vec{x})
\ee
 and we recognize the 3D special conformal transformations in flat $\mathbb{R}^3$.
We also  note that special conformal transformations can be written as 
\be
&&\frac{\tau'}{-\tau'^2+x'^2}=\frac{\tau}{-\tau^2+\vx^2},\\
&&\frac{x_i'}{-\tau'^2+x'^2}=\frac{x_i}{-\tau^2+\vx^2}+b_i \label{inv}
\ee
and therefore they can be generated by 
\be 
({\rm inversion})\times({\rm translation})\times({\rm inversion}),
\ee  
where by  inversion we mean
\be
&&\tau\to \tau'=\frac{\tau}{-\tau^2+\vx^2}, ~~~~x_i\to x_i'=\frac{x_i}{-\tau^2+\vx^2}. \label{inversion}
\ee
In other words, the special conformal transformations are generated by the transformation chain
\be
&\tau\to \tau'=\frac{\tau}{-\tau^2+\vx^2}, ~~&x_i\to x_i'=\frac{x_i}{-\tau^2+\vx^2}, ~~~ ~~(\mbox{inversion})\\
&\tau'\to \tau''=\tau', ~~&x_i'\to x_i''=x_i'+b_i,~~~ ~~~~(\mbox{translation})\\
&\tau''\to \tau'''=\frac{\tau''}{-\tau''^2+x''^2}, ~&x_i''\to x_i'''=\frac{x_i''}{-\tau''^2+x''^2},  ~(\mbox{inversion}).
\ee
Now, the action for a scalar field  
\be
{S}=\frac{1}{2}\int \de^4x\sqrt{-g}\left(-\p_\mu \phi\p^\mu\phi -m^2\phi^2\right) \label{scalar}
\ee
and on a de Sitter background 
it is written as 
\be
{S}=\frac{1}{2}\int \frac{\de^3x\de\tau}{H^2\tau^2}\left(\eta^{\mu\nu}\p_\mu\phi\p_\nu\phi-
\frac{m^2}{H^2\tau^2}\phi^2\right). \label{scalarex}
\ee
%In terms of the Fourier modes
%\be
%\phi(\vx,\tau)=\int\frac{\de^3k}{(2\pi)^{3/2}}e^{i\vec{k}\cdot \vec{x}}\phi_{\vk}(\t),
%\label{ffour}
%\ee
%the action (\ref{scalarex}) is expressed as
%\be
%S=\frac{1}{2}\int \frac{\de^3k\de\tau}{H^2\tau^2}\Big{\{}\phi_{\vk}'\phi_{-\vk}'+\left(k^2-\frac{m^2}{H^2\tau^2}\right)\phi_{\vk}\phi_{-\vk}
%\Big{\}}.
%\ee
It is easy to verify that (\ref{scalarex}) is invariant under the transformations (\ref{specconf}) if $\phi(\vx,\tau)$ transforms as
\be
\label{ff}
\phi(\vec{x},\t) \rightarrow \phi'(\vec{x},\t),
\ee
satisfying $
\phi'(\vec{x}',\tau')=\phi(\vec{x},\t)$.
Indeed, for an inversion 
\be
x^\mu\to {x^\mu}'=\frac{x^\mu}{x^2}
\ee
we have that 
\be
\p_\mu= (x^{'2})J^\nu_{\,\,\,\mu}\p_\nu',
\ee
where 
\be
J^\mu_{\,\,\,\nu}=\delta^\mu_{\,\,\,\nu}-2\frac{x^\mu x_\nu}{x^2}.
\ee
Then invariance of the action (\ref{scalarex}) follows from (\ref{ff}) and the relation
\be
J^\mu_{\,\,\,\nu} J^{\nu}_{\,\,\,\rho}=\delta^\mu_{\,\,\,\rho}. \label{jor}
\ee
The field equation for the scalar field is
\be
\p_{\t\t}^2\phi-\frac{2}{\tau}\p_\t\phi-\nabla^2\phi+\frac{m^2}{H^2\tau^2}\phi=0. \label{sca}
\ee
On super-Hubble scales we can isolate the time-dependent behavior as 
\be
\phi(\vx,\tau)\sim \tau^{\Delta_\pm}\chi(\vec{x}),   ~~~\Delta_\pm=\frac{3}{2}\left(1\pm\sqrt{1-\frac{4m^2}{9H^2}}\right). \label{deltascalar}
\ee
Let us therefore write for $H\tau\ll 1$ 
\be
\phi(\vx,\tau)=\left(H \tau \right)^\Delta \chi(\vec{x}),~~~~\p_\t\phi(\vx,\tau)=\frac{\Delta}{ \t}\phi(\vx,\tau),
\ee
where we have kept the dominant solution at large times, that is that with $\Delta=\Delta_-$. In this case we get 
\be
S=-\frac{1}{2}\int \frac{\de^3x\de\tau}{H^2\tau^2}\,\,
%\left(H \tau\right)^{2\Delta}
\left((\nabla \phi)^2-\frac{H^2\Delta^2-m^2}{H^2\tau^2}\phi^2\right). \label{scalarexd}
\ee
%Clearly, if we impose that the $\chi(\vx)$ field is a primary field under rescalings
%\be
%\vx\to \vx'=\lambda \vx, ~~~\tau\to \tau'=\lambda \tau, ~~~~\chi(\vx)\to \chi'(\vx')=\Omega_\lambda^{\Delta_\chi/2}\chi(\vx)=
%\lambda^{\Delta_\chi} \chi(\vx),
%\ee
%where $g'_{ij}(\vx')=\Omega_\lambda(\vx)g'_{ij}(\vx)$ sets the conformal factor under scalings, then 
%the action (\ref{scalarexd}) is invariant if
%\be 
%\label{condition}
%\Delta_\chi =-\Delta.
%\ee
Note that for $\tau^2\ll \vx^2$, the inversion (\ref{inversion}) is written as
\be
&&\tau\to \tau'=\frac{\tau}{|\vx|^2}, ~~~~x_i\to x_i'=\frac{x_i}{|\vx|^2}, 
\label{inver}
\ee
and leaves invariant the action (\ref{scalarexd}) as can easily be checked for a $\phi(\vx,\tau)$ satisfying
(\ref{ff}).  In addition, in this case, (\ref{scalarexd}) is also invariant under rescalings (\ref{scalings}). 
When the inversion (\ref{inver}) and the dilation (\ref{scalings}) are  combined with translations and
rotations, they generate the 3D conformal group acting on $\mathbb{R}^3$ and 
(\ref{sca})   therefore
possesses full 3D conformal invariance.
Note that the transformation of $\phi(\vx,\tau)$, see Eq.  (\ref{ff}),  is written in terms of $\chi$ as
\be
\phi'(\vx',\t')=(H\t')^\Delta \chi'(\vx')=\phi(\vx,\t)=(H\t)^{\Delta} \chi(\vx),
\ee
so that $\chi$ transforms under dilations (\ref{scalings}) and inversions (\ref{inver}) as
\be
\chi'(\lambda\vx)=\lambda^{-\Delta} \chi( \vx)\, , ~~~\chi'\left(\frac{\vx}{|\vx^2|}\right)=|\vx|^{2\Delta} \chi(\vx),
\ee
respectively. These are exactly the transformations we are expecting of a primary field of dimension $\Delta$. Therefore, 
on super-Hubble scales, the theory possesses 3D conformal invariance if the time-independent field $\chi(\vx)$ 
has scaling dimension $\Delta$. 

Let us rephrase in other terms and look at the infinitesimal version of the special conformal transformations  on super-Hubble scales, that is the transformation (\ref{asdf}). Under such an  infinitesimal (passive) transformation, the scalar field variation reads

\begin{eqnarray}
\delta\phi(\vx,\tau)&\simeq & 
\delta x^\mu\p_\mu\phi(\vx,\t) \nonumber\\
&=& -\delta\tau\partial_\tau \phi(\vx,\tau)+\delta\vx\cdot\nabla \phi(\vx,\tau)
=(2\vec{b}\cdot\vec{x})\tau\partial_\tau \phi(\vx,\tau)-(2\vec{b}\cdot\vec{x})\vx\cdot\nabla \phi(\vx,\tau)+\vx^2\,\vb\cdot\nabla \phi(\vx,\tau)\nonumber\\
&=&\Delta(2\vec{b}\cdot\vec{x}) \phi(\vx,\tau)-(2\vec{b}\cdot\vec{x})\vx\cdot\nabla \phi(\vx,\tau)+\vx^2\,\vb\cdot\nabla \phi(\vx,\tau),
\end{eqnarray}
which gives

\be
\delta\chi(\vx)=\Delta(2\vec{b}\cdot\vec{x}) \chi(\vx)-(2\vec{b}\cdot\vec{x})\vx\cdot\nabla \chi(\vx)+\vx^2\,\vb\cdot\nabla \chi(\vx).
\ee
This  is precisely the transformation of a primary field with weight $\Delta$ under 3D special conformal transformation. Therefore, correlation functions of scalar fields on de Sitter must be invariant
under conformal transformations of Euclidean three-space on the future boundary $\tau=0$. 
Similarly, an interaction of the form
\be
{S}_{\rm 3}=\int \de^4x\sqrt{-g}\mu\phi^3(\vx,\tau)=\int \frac{\de^3x \de\tau}{H^4\tau^4}\, \mu
\left(H\tau\right)^{3\Delta}\chi^3(\vx,\tau)  \label{int3}
\ee
is  invariant under 3D conformal symmetry. In fact, all non-derivative interactions of the form
\be
{S}_{\rm N}=\int \de^4x\sqrt{-g}\mu\phi^N=\int \frac{\de^3x \de\tau}{H^4\tau^4}\, \mu
\left(H\tau\right)^{N\Delta}\chi^N(\vx,\tau), 
\ee
are invariant under 3D conformal symmetry, although subleading with respect to (\ref{int3}) on super-Hubble scales for $N>3$.

\section{Conformal symmetries of de Sitter and the vector field}

%%%%%%%%%%%%%%%%%%%%%%%%%%%%%%%%%%%%%%%%%%%%%%%%%%%%%%%%%%%%%%%%%%%%%%%%%%%%%
%%%%%%%%%%%%%%%%%%%%%%%%%%%%%%%%%%%%%%%%%%%%%%%%%%%%%%%%%%%%%%%%%%%%%%%%%%%%%

\noindent
Let us consider now the case of a massless vector field $A_\mu(\vx,\t)$. In particular we consider the action
\be
S_{A}=-\frac{1}{4}\int \de^4x \sqrt{-g} I^2(\phi)F_{\mu\nu}F_{\kappa\rho}g^{\mu\kappa}g^{\nu\rho},
\ee
with gauge coupling $g=1/I(\phi)$. In the de Sitter background, this action is written as
\be
S_{A}=-\frac{1}{4}\int \de^3x\de\tau  I^2(\phi)F_{\mu\nu}F_{\kappa\rho}\eta^{\mu\kappa}\eta^{\nu\rho}, \label{sa}
\ee
where $\eta_{\mu\nu}$ is standard Minkowski metric.
In the $A_0=0$ gauge,  
the action (\ref{sa}) turns out to be
\be
S_{A}=-\frac{1}{4}\int \de^3x\de\tau  I^2(\phi)
\left({A_i'}^2-(\partial_i A_j-\partial_j A_i)^2\right). \label{sa1}
\ee
We will assume that 
\be
I(\phi)=I(\t)=(H\t)^{-n} \label{II}
\ee
and we would like to see for which values of $n$ the action
(\ref{sa1}) is invariant under the 3D conformal group. It is clear that the most stringent constraint 
will arise from the invariance under the inversion (\ref{inver}). 

We should recall that a vector $X_i$ 
with scaling dimension $\Delta_X$ in $D$-dimensions   transforms 
under the conformal group as $X_i(x)\to X_i'(x')$ where
\be
X_i'(x')=\left|\det \left( \frac{\partial {x'}^j}{\partial {x}^i}\right)\right|^{(1-\Delta_X)/D}
\frac{\partial x^j}{\partial {x'}^i}X_j(x). \label{con}
\ee
In general, a tensor $T_{i_1i_2\cdots i_n}$ transforms as 
\be
T'_{i_1i_2\cdots i_n}(x')=\left|\det \left( \frac{\partial {x'}^j}{\partial {x}^i}\right)\right|^{(n-\Delta_T)/D}
\left(\frac{\partial x^{j_1}}{\partial {x'}^{i_1}}\frac{\partial x^{j_2}}{\partial {x'}^{i_2}}
\cdots \frac{\partial x^{j_n}}{\partial {x'}^{i_n}}\right)T_{j_1j_2\cdots j_n}(x), \label{tensor}
\ee
where $\Delta_T$ is its conformal dimension.
Thus, for example, under rescalings we have
\be
X_i'(x')=\lambda^{-\Delta}\,  
X_i(x), \label{con1}
\ee
whereas for  the inversion (\ref{inver})
\be
X_i'(x')=|\vx|^{2+2(\Delta-1)}J^{\, j}_i(x) X_j(x).
\ee
The action (\ref{sa1}) is
\begin{align}
S_A=-\frac{1}{4}\int \de^3x\de\tau \left(H\t\right)^{-2n}
\Big{\{}(\partial_{\t}A_i)^2-(\partial_i A_j-\partial_j A_i)^2\Big{\}}. \label{sa3}
\end{align}
 Expressing $A_i$ in terms  of the field $a_i$  as
\be
A_i(\vx,\t)=(H\t)^n\, a_i(\vx,\t),  \label{Aa}
\ee
we get that 
\be
S_A=-\frac{1}{4}\int \de^3x\de\tau 
\left\{(\partial_{\t}a_i)^2+\frac{n(n+1)}{\t^2}a_i^2-(\partial_i a_j-\partial_j a_i)^2\right\}. \label{sa4}
\ee
It can be checked that the action (\ref{sa4}) can only be invariant under inversion (\ref{inver}) if 
the dimension of the vector $a_i$ is $\Delta_a=1$. Indeed, 
under inversions, the vector $a_i$ transforms as
\be
a_i'(x')=\left|\det \left( \frac{\partial {x'}^j}{\partial {x}^i}\right)\right|^{(1-\Delta_a)/D}  
\frac{\partial x^j}{\partial {x'}^i}\, a_j(x). \label{atr}
\ee
This is a coordinate transformation augmented by the factor 
$J^{(1-\Delta_a)/D}$ where $J=\Big{|}\det(\partial {x'}^j/\partial {x}^i)\Big{|}$. Then, 
when transforming $f_{ij}=(\partial_i a_j-\partial_j a_i)$, there will appear cross terms of the 
form $a_j\partial_iJ$ which cannot be canceled and will spoil conformal invariance. In fact, conformal invariance 
can be maintained if the 
$J$ factor in the transformation (\ref{atr}) is missing. This is possible for
\be
\Delta_a=1.
\ee
Then, by using that 
\be
f'_{ij}=|\vx|^4\, J_{ik}\, J_{jl}\, f_{kl}\, , ~~~~\partial_{\t'} a_i'=|\vx|^4 \,J_{ij}\, \partial_{\t} a_j \label{ftr}
\ee
 and the orthogonality relation (\ref{jor}), the action (\ref{sa4}) turns out to be under the inversion (\ref{inver}) 
\begin{align}
S_{A'}&=-\frac{1}{4}\int \de^3x'\de\tau' 
\left\{(\partial_{\t'}a_i')^2+\frac{n(n+1)}{{\t'}^2}{a'_i}^2-(\partial'_i a'_j-\partial'_j a'_i)^2\right\}\nonumber \\
&=-\frac{1}{4}\int \frac{\de^3x\de\tau}{|\vx|^8}  |\vx|^{8}
\left\{(\partial_{\t}a_i)^2+\frac{n(n+1)}{\t^2}a_i^2-(\partial_i a_j-\partial_j a_i)^2\right\}\nonumber\\
&=S_A&. \label{sa3'}
\end{align}
As we did for the scalar, the field equations for $a_i$ at super-Hubble scales is 
\be
\partial_{\tau\tau}^2a_i-\frac{n(n+1)}{\t^2}a_i=0.  \label{eqa}
\ee
The general solution is written as 
%the solutions of which is 
\be
a_i=(H\t)^{-n} V_i(\vx)+(H\t)^{1+n}U_i(\vx).  \label{sola}
\ee
Depending on the value of $n$, the leading term is different. So, there will be two cases, which we will call ``magnetic''
and ``electric'' respectively. 
\vskip.2in

%%%%%%%%%%%%%%%%%%%%%%%%%%%%%%%%%%%%%%%%%%%%%%%%%%%%%%%%%%%%%%%%%%%%%%%%%%%%%

\subsection{The magnetic case}

%%%%%%%%%%%%%%%%%%%%%%%%%%%%%%%%%%%%%%%%%%%%%%%%%%%%%%%%%%%%%%%%%%%%%%%%%%%%%

The excitations of the vector field during a de Sitter epoch generate magnetic-like fluctuations if  $n>-1/2$.
In this case, the solution is 
\be
a_i=(H\t)^{-n} V_i(\vx) \label{Va}
\ee
and   the conformal  dimension of $V_i$ can be determined by the conformal dimension of $a_i$ (which is $\Delta_a=1$) 
to be
\be
\Delta_V=-n+1. \label{v1}
\ee
Indeed, under rescaling $\t\to\t'=\lambda \t$ and $\vx\to \vx'=\lambda \vx$ we get
\be
a_i'(\vx',\t')=(H\t')^{-n}V_i'(\vx')=\lambda^{-n}(H\t)^{-n}V_i'(\vx')=\lambda^{-1}a_i(\vx,\t)=\lambda^{-1}(H\t)^{-n}V_i(\vx)\
\ee
and therefore,
\be
V_i'(\vx')=\lambda^{n-1}V_i(\vx).
\ee
Then, a simple comparison with (\ref{con1}) reveals that $V_i$ transforms as a vector of conformal dimension $(-n+1)$, that is Eq. (\ref{v1}).  
Let us note that the electric and magnetic  fields are given by 
\be
E_i=-\frac{I}{a^2}{A_i}'\, , ~~~B_i=\frac{I}{a^2}\epsilon_{ijk}\partial_j A_{\vk}.
\ee
Then, by using (\ref{Aa}) and (\ref{Va}) we get 
\be
E_i=-\frac{I}{a^2}{V_i}'=0\, , ~~~B_i=\frac{I}{a^2}\epsilon_{ijk}\partial_j V_{\vk}.
\ee
Therefore, in the magnetic case, the electric field vanish and the magnetic field is constant on super-Hubble scales only for 
\be
I=a^2\, ,\
\ee
that is only for  $n=2$. 

%%%%%%%%%%%%%%%%%%%%%%%%%%%%%%%%%%%%%%%%%%%%%%%%%%%%%%%%%%%%%%%%%%%%%%%%%%%%%

\subsection{The electric  case}

%%%%%%%%%%%%%%%%%%%%%%%%%%%%%%%%%%%%%%%%%%%%%%%%%%%%%%%%%%%%%%%%%%%%%%%%%%%%%

The excitations of the vector field during a de Sitter epoch generate electric-like fluctuations if  $n<-1/2$.
In such a case, the solution is 
\be
a_i=(H\t)^{n+1} U_i(\vx) \label{Ua}
\ee
and the conformal dimension of $U_i$ is 
\be
\Delta_U=n+2. \label{Uc}
\ee
Indeed, under rescalings
\be
\label{asd}
a_i'(\vx',\t')=(H\t')^{n+1}U_i'(\vx')=\lambda^{n+1}(H\t)U_i'(\vx')=\lambda^{-1}a_i(\vx,\t)=
\lambda^{-1}(H\t)^{n+1}U_i(\vx)
\ee
and (\ref{Uc}) easily follows. 
Note also that in this case, by using (\ref{Aa}) and (\ref{Ua}) the electric and magnetic fields are
\be
&&E_i=-\frac{I}{a^2}(2n+1)H(H\t)^{2n}{U_i}=-(2n+1)H(H\t)^{n+2}U_i\, , \nonumber \\
&&B_i=\frac{I}{a^2}(H\t)^{2n+1}\epsilon_{ijk}\partial_j U_{\vk}=(H\t)^{n+3}\epsilon_{ijk}\partial_j U_{\vk}.
\ee
Therefore, in the electric case, the magnetic field vanish on super-Hubble scales whereas the electric field is constant
only for  
\be
I=a^{-2}\, ,
\ee
i.e., for $n=-2$. Of course, this is related to the electric-magnetic duality in this case since
under $I\to 1/I$, electric and magnetic fields exchange their role\cite{peloso}.

\section{The correlators of the vector field}

%%%%%%%%%%%%%%%%%%%%%%%%%%%%%%%%%%%%%%%%%%%%%%%%%%%%%%%%%%%%%%%%%%%%%%%%%%%%%

In order to describe the higher-order correlator of the vector fields we need to specify the dependence of the gauge coupling to the inflaton field, that is the function $I(\phi)$. 
We consider $I(\tau)=I(\phi_0(\t))$ as   $I(\phi)$ at the background value of $\phi=\phi_0(\tau)$ during inflation. Fluctuations 
of the latter produce interactions of the form
\be
S_{\rm int}=\int \de^3x \de\tau \, \,  \left.\frac{\p I^2}{\p \phi}\right|_{\phi_0}\delta\phi 
\left({A_i}'{A_i}'-(\nabla A_i)^2\right).\label{vecper}
\ee
Following Ref. \cite{I1}
for $
I(\phi)=a^{n}(\tau)$, we  should have
\be
I^2=\exp\left(-2 n \int^\phi\frac{V(\phi')}{V'(\phi')} \de\phi'\right).
\ee
Then, we easily find that 
\be
\frac{\p I^2}{\p \phi}\Big{|}_{\phi_0}=-2n\, I^2(\phi_0)\frac{V(\phi_0)}{V'(\phi_0)}.
\ee
By combining Friedmann and inflaton equations we find that 
\be
\frac{V(\phi_0)}{V'(\phi_0)}=-\frac{\de\ln a}{\de\phi}
\ee
and since 
\be
a=\left(\frac{\tau}{\tau_0}\right)^{1+b}, ~~~\phi_0=v+\sqrt{2\epsilon}(1+b)\ln \tau,  \label{bac}
\ee
where $\epsilon$ is the slow roll parameter, we get that 
\be
\frac{\de\ln a}{\de\phi}=\frac{1}{\sqrt{2\epsilon}}.
\ee
Therefore, we find 
\be
\frac{\p I^2}{\p \phi}\Big{|}_{\phi_0}=\sqrt{\frac{2}{\epsilon}}\, n \,I^2(\phi_0)\, . \label{if}
\ee
The above equation (\ref{if}) can be integrated to give
\be
I^2(\phi)=I_0^2 e^{\sqrt{\frac{2}{\epsilon}}\, n \, (\phi-\phi_0)}
\ee
Therefore the action (\ref{sa1}) turns out to be
\be
S_{A}=-\frac{1}{4}\int \de^3x\de\tau I_0^2 e^{\sqrt{\frac{2}{\epsilon}}\, n\, (\phi-\phi_0)}
\left({A_i'}^2-(\partial_i A_j-\partial_j A_i)^2\right), \label{sa11'}
\ee
Expanding around $\phi_0$ given by the relation  (\ref{bac}), the action turns out to be
\be
S_{A}=-\frac{1}{4}\int \de^3x\de\tau  (H\t)^{-2n}\left(1+\sqrt{\frac{2}{\epsilon}}\, n \,\delta\phi+
\frac{n^2}{\epsilon}
\delta\phi^2+\cdots\right)
\left({A_i'}^2-(\partial_i A_j-\partial_j A_i)^2\right).\label{sa11}
\ee
Clearly, the interactions in the action (\ref{sa11}) are conformal invariant if $\phi(\vx,\t)$ has scaling dimension zero
$\Delta_\phi=0$ which we will assume from now on (therefore  neglecting the small deviations from scale-invariance). In other words,
the interaction of the vector field with the inflaton field are conformal invariant at any order in perturbation theory. This will become important in the following. We are now in the position to characterize the correlators involving the vector field.

Let us first recall that in momentum space that the Ward identities associated to 
dilations and special conformal transformations in general $D$-dimensions of
a symmetric, two-tensor $N$-point amplitude ${\cal{A}}'_{lm}$ (primes indicate they are computed without 
Dirac delta functions), using (\ref{tensor}),  are given by 
\begin{align}
 \delta_\lambda {\mathcal{A}}'_{lm}&=\left\{-d(N-1)+\sum_{a=1}^N
 \left(\Delta_a-\vec{k}_a\cdot\vec{\partial}_{k_a}\right)\right\}{\mathcal{A}}'_{lm} \label{D},\\
 \delta_{b^i}{\mathcal{A}}'_{lm}&
 =i\sum_{a=1}^N\left\{2(\Delta_a-d)\partial_{k_a^i}+k_a^i\nabla^2_{k_a}-2\vec{k}_a \cdot
\vec{\partial}_{k_a}\partial_{k_a^i}
%\right.\nonumber \\
% &+2 (\delta^i_n\partial_{k_a}        \left.
 \right\}{\mathcal{A}}'_{lm}\nonumber \\
 &-2i\sum_{a=1}^N\left\{\left(\delta^{ni}\partial_{k_a^l}-\delta_l^i
 \partial_{k_a^n}\right){\mathcal{A}}'_{nm}+\left(\delta^{ni}\partial_{k_a^m}-\delta_m^i
 \partial_{k_a^n}\right){\mathcal{A}}'_{ln}\right\}.\label{K}
\end{align}
We will see that the above transformations, together with the fact that ${\cal{A}}'_{lm}$ transforms as a symmetric
two-tensor under SO(3) rotations is enough to determine their form.  

\subsection{The magnetic case}
In the temporal $A_0=0$ gauge, the  field  $A_i$ as well as $a_i$ and $V_i$ are divergenceless  
$\nabla^i A_i=\nabla^ia_i=\nabla^iV_i=0$.
Then, by SO(3) covariance and momentum conservation, we get
\be
\lg A_i(\vk)A_j(-\vk)\rg'=P(k)\left(\delta_{ij}-\alpha(k)k_ik_j\right)
\ee
for the Fourier modes $A_i(\vk)$ of the vector $A_i(\vx)$. The divergenceless condition for $A_i(\vx)$ is written as 
$k_i\lg A_i(\vk)A_j(-\vk)\rg'=0$, which specifies $\alpha=1/k^2$ so that 
\be
\lg A_i(\vk)A_j(-\vk)\rg'=P(k)\left(\delta_{ij}-\frac{k_ik_j}{k^2}\right). \label{2pt-m}
\ee
Finally, by using that 
\be
\vk\cdot\vec{\partial}_{\vk}\left(\delta_{ij}-\frac{k_ik_j}{k^2}\right)=0,
\ee
$P(k)$ is specified by the invariance under dilations and special conformal transformations (\ref{D}) and (\ref{K})
to satisfy
\be
\Big{\{}\!-3+2(-n+1)-k\partial_{\vk}\Big{\}}P(k)=0.
\ee
The SO(3) symmetric solution to the above equation is 
\be
P(k)=\frac{C_M}{k^{2n+1}},
\ee
where $C_M$ is a constant. 
We can similarly determine  the three-point function $\lg \delta\phi A_i A_j\rg$ by rotational and 3D conformal symmetry.
%\begin{align}
%\lg \delta \phi(\vx) a_i(\vy)a_j(\vec{z})\rg&=
%\sum_{\lambda,\lambda'}
% \int\frac{\de^3k_1\de^3k_2\de^3k_3}{(2\pi)^{9/2}}e^{i(\vec{k_1}\cdot \vec{x}+\vec{k_2}\cdot \vec{y}+\vec{k_3}\cdot \vec{z})}
%\nonumber \\
%&\delta(\vec{k_1}+\vec{k}_2+\vec{k}_3)e^\lambda_{i}(\vec{k_2})e^{\lambda'}_{j}(\vec{k_3})\lg \delta\phi_{\vk_1}
%u^\lambda_{\vec{k_2}}u^{\lambda'}_{\vec{k_3}}\rg,
%\end{align}
%where we have used (\ref{ffour}) and the decomposition of $\delta\phi$ in Fourier modes $\delta\phi_{\vk}$. 
%As a result, we get that 
%\be
%\lg\delta\phi(\vk_1) a_i(\vk_2)a_j(\vk_3)\rg=\sum_{\lambda,\lambda'}e^{\lambda}_{i}(\vec{k_2})e^{\lambda'}_{j}(\vec{k_3})
%\lg \delta\phi_{k_1}u^\lambda_{\vec{k_2}}u^{\lambda'}_{\vec{k_3}}\rg.
%\ee
SO(3) covariance imposes for the Fourier modes $A_i(\vk)$ of $A_i(\vx)$ the form

\begin{align}
\lg\delta\phi(\vk_1) A_i(\vk_2)A_j(\vk_3)\rg ' &
%=\sum_{\lambda,\lambda'}e^{\lambda}_{i}(\vec{k_2})e^{\lambda'}_{j}(\vec{k_3})
%\lg \delta\phi_{k_1}u^\lambda_{\vec{k_2}}u^{\lambda'}_{\vec{k_3}}\rg=\nonumber \\
%&
=c_1\, \,\delta_{ij}+c_2\, \, (k_2)_i(k_3)_j+c_3\, \, (k_2)_j(k_3)_i+c_4\,\, (k_2)_i(k_2)_j+c_5 \,\, (k_3)_j(k_3)_i,
\end{align}
where $c_i=c_i(\vec{k}_1,\vec{k}_2,\vec{k}_3)$. 
By multiplying by $(k_2)^i$  and by $(k_3)^j$, and using that $A_i$ is divergenceless ($\vk^i\cdot A_i(\vk)=0$),
we get the conditions

\begin{align}
 0&=c_1 (k_2)_j+c_2 k_2^2(k_3)_j+c_3(\vec{k}_2\cdot\vec{k}_3)(k_2)_j+c_4k_2^2 (k_2)_j+c_5(\vec{k}_2\cdot\vec{k}_3)(k_3)_j\nonumber\\
 0&=c_1 (k_3)_i+c_2 k_3^2(k_2)_i+c_3(\vec{k}_2\cdot\vec{k}_3)(k_3)_i+c_4(\vec{k}_2\cdot\vec{k}_3) (k_2)_i+
 c_5k_3^2(k_3)_i.
\end{align}
The above equations specify the constants as 
\be
c_3=\frac{c_2 k_2^2 k_3^2}{(\vec{k}_2\cdot\vec{k}_3)^2}-\frac{c_1}{(\vec{k}_2\cdot\vec{k}_3)}\, , ~~~
c_4=-\frac{c_2 k_3^2}{(\vec{k}_2\cdot\vec{k}_3)}\, , ~~~c_5=-\frac{c_2 k_2^2}{(\vec{k}_2\cdot\vec{k}_3)}\, , \label{tr}
\ee
and therefore, by appropriate parametrization, the three-point correlator can be written as
\begin{align}
&\lg\delta\phi(\vk_1) a_i(\vk_2)a_j(\vk_3)\rg'= I_1\, D_{ij}+I_2\,  \Delta_{ij}, \label{cor-I}
%(k_2)_i(k_3)_j+\frac{k_2^2k_3^2(k_2)_j(k_3)_i}{(\vec{k}_2\cdot\vec{k}_3)^2}-\frac{(k_2)_i(k_2)_jk_3^2}{
%\vec{k}_2\cdot\vec{k}_3}-\frac{(k_3)_i(k_3)_jk_2^2}{
%\vec{k}_2\cdot\vec{k}_3}\right)
\end{align}
where $I_{1,2}=I_{1,2}(\vk_1,\vk_2,\vk_3)$ and 
\be
\Delta_{ij}=\vec{k}_2\cdot\vec{k}_3\delta_{ij}-(k_2)_j(k_3)_i\, , ~~~
D_{ij}=\delta_{ij}-\frac{(k_2)_i(k_2)_j}{k_2^2}-\frac{(k_3)_i(k_3)_j}{k_3^2}+\frac{(k_2)_i(k_3)_j}{k_2^2k_3^2}\vec{k}_2
\cdot\vec{k}_3.
\ee
%where we have parametrized 
%\be
%c_1=  I_2\, (\vec{k}_2\cdot\vec{k}_3)\,  , ~~~~c_2=I_1
%\ee
%and $I_2=I_2(\vec{k}_1,\vec{k}_2,\vec{k}_3),\, I_1=I_1(\vec{k}_1,\vec{k}_2,\vec{k}_3)$. 
The final step is to implement conformal invariance.  
Since  the two terms in (\ref{tr}) containing $I_1$ and $I_2$ respectively are independent,
they should be conformal invariant independently. 
%At this point let us recall that in momentum space
%dilations and special conformal transformations in general $d$ dimensions of $N$-point amplitudes ${\cal{A}}'_{lm}$ (primes indicate they are computed without 
%$\delta$ functions)  are defined as 
%\begin{align}
% \delta_D {\mathcal{A}}'_{lm}&=\left\{-d(N-1)+\sum_{a=1}^N
% \left(\Delta_a-\vec{k}_a\cdot\vec{\partial}_{k_a}\right)\right\}{\mathcal{A}}'_{lm} \label{D},\\
% \delta_{K^i}{\mathcal{A}}'_{lm}&
% =i\sum_{a=1}^N\left\{2(\Delta_a-d)\partial^i_{k_a}+k_a^i\nabla^2_{k_a}-2\vec{k}_a
%\vec{\partial}_{k_a}\partial^i_{k_a}
%\right.\nonumber \\
% &+2 (\delta^i_n\partial_{k_a}        \left.
% \right\}{\mathcal{A}}'_{lm}\nonumber \\
% &-2i\sum_{a=1}^N\left\{\left(\delta^{ni}\partial_{k_a^l}-\delta_l^i
% \partial_{k_a}^n\right){\mathcal{A}}'_{nm}-\left(\delta^{ni}\partial_{k_a^m}-\delta_m^i
% \partial_{k_a}^n\right){\mathcal{A}}'_{ln}\right\}.\label{K}
%\end{align}
On super-Hubble scales we may assume that $I_{1,2}$ have a  series expansion in terms of $k_1,k_2,k_3$ as 
\be
I_1=\alpha_M k_1^{a_1}(k_2 k_3)^{b_1}+\ldots \, , ~~~I_2=\beta_M k_1^{a_2}(k_2 k_3)^{b_2}+\cdots, \label{i1i2}
\ee
where $\alpha_1,\alpha_2$ are constants and  we have also used the symmetry of the correlator (\ref{cor-I}) 
under simultaneous interchange $i\leftrightarrow 
j$ and $\vk_2\leftrightarrow \vk_3$. 
By using the relations 
\be
&&\vk_2\cdot\vec{\partial}_{k_2}\Delta_{ij}=\vk_3\cdot\vec{\partial}_{k_3}\Delta_{ij}=\Delta_{ij},\nonumber \\
&&\vk_2\cdot\vec{\partial}_{k_2}D_{ij}=\vk_3\cdot\vec{\partial}_{k_3}D_{ij}=0,
\ee
and the fact the scaling dimensions of $\delta\phi$ and $a_i$ are $\Delta_{\delta\phi}=0$ and $\Delta_A=\Delta_V=-n+1$, 
respectively,
we get from  scaling invariance (\ref{D}) that that 
%$a_1=a_2=0~
$b_1=-n-2-a_1/2,~ b_2=-n-3-a_2/2$, i.e., 
\be
I_1=\frac{\alpha_M\, k_1^{a_1}}{(k_2k_3)^{n+2+\frac{a_1}{2}}}\, , ~~~I_2=\frac{\beta_M\, k_1^{a_2}}{(k_2k_3)^{n+3+\frac{a_2}{2}}}.
\ee
In addition, the  implementation of special conformal invariance (\ref{K}) gives that $a_2=0$ and $\alpha_M=0$ 
since it turns out that only $I_2$ is invariant (up to
${\cal{O}}\big{(}(k_2k_3)^{-n-2}\big{)}$ terms). 
Thus, conformal invariance with constant magnetic field at super Hubble scales ({\it i.e.} $n=2$),
specify the three-point function to be
\be
\fbox{$\displaystyle
\lg\delta\phi(\vk_1) A_i(\vk_2)A_j(\vk_3)\rg' =
%\frac{\alpha_M}{k_2^4k_3^4}\left(
%\delta_{ij}-\frac{(k_2)_i(k_2)_j}{k_2^2}-\frac{(k_3)_i(k_3)_j}{k_3^2}+\frac{(k_2)_i(k_3)_j}{k_2^2k_3^2}\vec{k}_2
%\cdot\vec{k}_3\right)\nonumber \\&+
\frac{\beta_M}{k_2^5k_3^5}\left(\vec{k}_2\cdot\vec{k}_3\delta_{ij}-(k_2)_j(k_3)_i\right). $}\label{3magn}
\ee
We should mention that we could have consider instead of (\ref{i1i2}), the most general form
\be
I_1=\alpha_M k_1^{a_1}(k_2 k_3)^{b_1}(\vk_2\cdot\vk_3)^{q_1}+\ldots \, , ~~~
I_2=\beta_M k_1^{a_2}(k_2 k_3)^{b_2}(\vk_2\cdot\vk_3)^{q_2}+\cdots, \label{i3}
\ee
However, although scaling symmetry (\ref{D}) can be satisfied with
\be
I_1=\frac{\alpha_M\, k_1^{a_1}(\vk_2\cdot\vk_3)^{q_1}}{(k_2k_3)^{n+2+q_1+\frac{a_1}{2}}}\, , ~~~
I_2=\frac{\beta_M\, k_1^{a_2}(\vk_2\cdot\vk_3)^{q_2}}{(k_2k_3)^{n+3+q_2+\frac{a_2}{2}}},
\ee
special conformal symmetry leads again to $q_2=0,\, a_2=0$ and $\alpha_M=0$,  that is again to the solution (\ref{3magn}).

\subsection{The electric case} 
For the case of electric-like excitations, 
again by  SO(3) covariance and momentum conservation, we get
\be
\lg U_i(\vk)U_j(-\vk)\rg'=P(k)\left(\delta_{ij}-\frac{k_ik_j}{k^2}\right)
\ee
and $P(k)$ is similarly  specified by the invariance under dilations and special conformal transformations (\ref{D}) and (\ref{K})
to satisfy
\be
\Big{\{}\!-3+2(n+2)-k\partial_{\vk}\Big{\}}P(k)=0.
\ee
The SO(3) symmetric solution to the above equation is 
\be
P(k)=C_E\, k^{2n+1},
\ee
where $C_E$ is a constant. Then, since $A_i=(H\t)^{2n+1} U_i$, we get 
\be
\lg A_i(\vk)A_j(-\vk)\rg'=(H\t)^{4n+2}C_E\, k^{2n+1}\left(\delta_{ij}-\frac{k_ik_j}{k^2}\right).
\ee
In particular, for a constant electric field at super-Hubble scales, $n=-2$ and we get
\be
\lg A_i(\vk)A_j(-\vk)\rg'=(H\t)^{-6}\frac{C_E}{k^3}\left(\delta_{ij}-\frac{k_ik_j}{k^2}\right),
\ee
whereas, the two-point correlator of the Fourier modes of the electric field is
\be
\lg E_i(\vk)E_j(-\vk)\rg'=9H^2\frac{C_E}{k^3}\left(\delta_{ij}-\frac{k_ik_j}{k^2}\right).
\ee
Similarly, it is straightforward to specify the three-point correlator 
$\lg \delta\phi A_i A_j\rg$ by rotational and 3D conformal symmetry.
In this case we get 
\begin{align}
&\lg\delta\phi(\vk_1) U_i(\vk_2)U_j(\vk_3)\rg' = I_1\, D_{ij}+I_2\,  \Delta_{ij}, \label{cor-II}
\end{align}
where $I_{1,2}=I_{1,2}(\vk_1,\vk_2,\vk_3)$. Again, we may assume that $I_{1,2}$ have a  series expansion in terms of 
$k_1,k_2,k_3$ at super-Hubble scales as 
\be
I_1=\alpha_E k_1^{a_1}(k_2 k_3)^{b_1}+\cdots \, , ~~~I_2=\beta_E k_1^{a_2}(k_2 k_3)^{b_2}+\cdots.
\ee
Then, by recalling that  the scaling dimensions of $\delta\phi$ and $a_i$ are $\Delta_{\delta\phi}=0$ and 
$\Delta_U=n+2$, 
respectively, scaling invariance (\ref{D})  of the three-point correlator gives
that 
%$a_1=a_2=0,~
$b_1=n-1-a_1/2,~ b_2=n-2-a_2/2$, that is
\be
I_1=\frac{\alpha_E\, k_1^{a_1}}{(k_2k_3)^{1-n+\frac{a_1}{2}}}\, , 
~~~I_2=\frac{\beta_E\, k_1^{a_2}}{(k_2k_3)^{2-n+\frac{a_2}{2}}}.
\ee
In addition, by employing  special conformal invariance we get now $a_1=0$ and $\beta_E=0$ 
since in this case only $I_1$ is invariant (up to  order 
${\cal{O}}\big{(}(k_2k_3)^{n-1}\big{)}$ terms) and hence
\begin{align}
\lg\delta\phi(\vk_1) U_i(\vk_2)U_j(\vk_3)\rg'&= \frac{\alpha_E}{(k_2k_3)^{1-n}}\left(
\delta_{ij}-\frac{(k_2)_i(k_2)_j}{k_2^2}-\frac{(k_3)_i(k_3)_j}{k_3^2}+\frac{(k_2)_i(k_3)_j}{k_2^2k_3^2}\vec{k}_2
\cdot\vec{k}_3\right).
%\nonumber \\
%&+\frac{\alpha_2}{(k_2k_3)^{2-n}}\left(\vec{k}_2\cdot\vec{k}_3\delta_{ij}-(k_2)_j(k_3)_i\right).
\end{align}
For a  constant electric field at super Hubble scales $n=-2$, we therefore get 
\begin{align}
\lg\delta\phi(\vk_1) A_i(\vk_2)A_j(\vk_3)\rg'&= (H\t)^{-6}\frac{\alpha_E}{(k_2k_3)^{3}}\left(
\delta_{ij}-\frac{(k_2)_i(k_2)_j}{k_2^2}-\frac{(k_3)_i(k_3)_j}{k_3^2}+\frac{(k_2)_i(k_3)_j}{k_2^2k_3^2}\vec{k}_2
\cdot\vec{k}_3\right) \label{3ele}
%\nonumber \\
%&+(H\t)^{-6}\frac{\beta_E}{(k_2k_3)^{4}}\left(\vec{k}_2\cdot\vec{k}_3\delta_{ij}-(k_2)_j(k_3)_i\right)
\end{align}
and 
\be
\label{3elec}
\fbox{$\displaystyle
\lg\delta\phi(\vk_1) E_i(\vk_2)E_j(\vk_3)\rg'= \frac{9H^2\, \alpha_E}{(k_2k_3)^{3}}\left(
\delta_{ij}-\frac{(k_2)_i(k_2)_j}{k_2^2}-\frac{(k_3)_i(k_3)_j}{k_3^2}+\frac{(k_2)_i(k_3)_j}{k_2^2k_3^2}\vec{k}_2
\cdot\vec{k}_3\right).$}
%\nonumber \\
%&\frac{9H^2\,\beta_E}{(k_2k_3)^{4}}\left(\vec{k}_2\cdot\vec{k}_3\delta_{ij}-(k_2)_j(k_3)_i\right).
\ee
Note that again, introducing powers  $(\vk_2\cdot\vk_3)^q$ in $I_1,\,I_2$ as we did in the magnetic case, 
do not change the 
(\ref{3ele}) as special conformal invariance (\ref{K}) leads again to $q=0$.  

The results (\ref{3magn}) and (\ref{3elec}) are fully  dictated by dilation and special conformal invariance. Indeed, since 
the de Sitter isometry group acts  as conformal group 
on $\mathbb{R}^3$ when the fluctuations are super-Hubble, the correlators must be of the form we found when perturbations
are on super-Hubble scales. In other words, correlators may have a more complicated form in momentum space. Nevertheless, as soon as they are evaluated with all the corresponding wavelengths outside the Hubble radius, the symmetries of the problem dictate their form.
This simple argument imposes that, while the amplitude of the correlators is not fixed by the symmetries, for sure it has to depend on a parameter signaling how long modes live outside the Hubble radius till the end of inflation. An educated guess for such a parameter is
the number of e-folds $N(k_t)=\ln(-k_t\tau)$ the given mode $k_t=(k_1+k_2+k_3)$ spends outside the Hubble radius.

This expectation is nicely confirmed by  comparing our results with those of Refs. \cite{motta,sloth,peloso} where the three-point correlator among the
inflaton field and two vector fields have been computed. A  simple inspection of their results, {\it e.g.} Eq. (4.16) of Ref. \cite{sloth} for the case
$n=2$ or Eqs. (46) and (47) of Ref. \cite{motta} for the case $n=-2$ (once some errors on the tensor parts are properly corrected), reveal that, despite the complexity of the full result,  on super-Hubble scales there is only one  dominant piece which is indeed  
 proportional to  $N(k_t)=\ln(-k_t\tau)$. 
The reason is simple: 
as soon as $N(k_t)$ gets larger than unity, the symmetry arguments apply and the correlator has to acquire the shape dictated by the symmetries at hand.

The symmetry arguments are in fact even more powerful. In the case in which there exists a background electric field (possibly sustained by IR super-Hubble modes), but whose associated energy density is negligible, the 3D conformal symmetry is 
not spontaneously broken. This is as a consequence of the fact that the electric field $\vec{E}$, or equivalently the field $\vec{U}$  in  
Eq. (\ref{asd}) (corresponding to $n=-2$), has zero conformal weight.  
This implies that the shape of correlators are the same  at any order in perturbation theory. 
This is indeed what has been found in Ref. \cite{peloso} 
(once the super-Hubble modes contributing to the electric field renormalize the background value), 
where the  one-loop corrections  for the curvature power spectrum and bispectrum have 
been shown to be the same as the tree-level ones.

\section{Conclusions}

%%%%%%%%%%%%%%%%%%%%%%%%%%%%%%%%%%%%%%%%%%%%%%%%%%%%%%%%%%%%%%%%%%%%%%%%%%%%%
%%%%%%%%%%%%%%%%%%%%%%%%%%%%%%%%%%%%%%%%%%%%%%%%%%%%%%%%%%%%%%%%%%%%%%%%%%%%%

In this paper we have investigated the properties of the vector perturbations generated during a primordial period of inflation
on the basis of the symmetries present during such a stage. The key point of all our logic is that the de Sitter isometry group acts  as 
conformal group  on the three-dimensional Euclidean space  for  the super-Hubble fluctuations. This allows to   
characterize the correlators involving the inflaton and the vector fields,  determining their shapes and shedding some light on some results found in the recent literature. Our results may be relevant for the following reason: when analyzing the broken statistical invariances of the CMB modes, one needs some sort of guidance to parametrize such deviations from the standard set-up. Symmetries may provide such a guidance in the very same way the do when writing effective field theories in the infrared because the ultraviolet completion is missing. We will investigate this issue in a separate publication.

 \section*{Acknowledgments}
A.R. and H.P. are  supported by the Swiss National
Science Foundation (SNSF), project `The non-Gaussian Universe" (project number: 200021140236). M.B. is supported by the Swiss National Science Foundation (SNSF).

%%%%%%%%%%%%%%%%%%%%%%%%%%%%%%%%%%%%%%%%%%%%%%%%%%%%%%%%%%%%%%%%%%%%%%%%

%%%%%%%%%%%%%%%%%%%%%%%%%%%%%%%%%%%%%%%%%%%%%%%%%%%%%%%%%%%%%%%%%%%%%%%

\end{document}